\documentclass{sig-alternate-05-2015}
\usepackage{color} 
\usepackage{booktabs} 
\usepackage{algorithm2e}
\usepackage{amsmath}
\usepackage{amssymb}
\usepackage{url}

\usepackage{subcaption}
\usepackage{cleveref}
\usepackage{graphicx}

\newdef{definition}{Definition}

\makeatletter
\def\@copyrightspace{\relax}
\makeatother

\begin{document}


\title{Around Average Behavior: 3-lambda Network Model} 

\numberofauthors{3}
\author{
%
%
\alignauthor
Milos Kudelka\\
       \affaddr{FEI, VSB - Technical University of Ostrava}\\
       \affaddr{17. listopadu 15, 708 33}\\
       \affaddr{Ostrava - Poruba, Czech Republic}\\
       \email{milos.kudelka@vsb.cz}
\alignauthor
Eliska Ochodkova\\
       \affaddr{FEI, VSB - Technical University of Ostrava}\\
       \affaddr{17. listopadu 15, 708 33}\\
       \affaddr{Ostrava - Poruba, Czech Republic}\\
       \email{eliska.ochodkova@vsb.cz}
\alignauthor
Sarka Zehnalova\\ 
       \affaddr{FEI, VSB - Technical University of Ostrava}\\
       \affaddr{17. listopadu 15, 708 33}\\
       \affaddr{Ostrava - Poruba, Czech Republic}\\
       \email{sarka.zehnalova@vsb.cz}
}
	   
\date{19 October 2016}

\maketitle
\begin{abstract}
The analysis of networks affects the research of many real phenomena. The complex network structure can be viewed as a network's state at the time of the analysis or as a result of the process through which the network arises. Research activities focus on both and, thanks to them, we know not only many measurable properties of networks but also the essence of some phenomena that occur during the evolution of networks. One typical research area is the analysis of co-authorship networks and their evolution. In our paper, the analysis of one real-world co-authorship network and  inspiration from existing models form the basis of the hypothesis from which we derive new 3-lambda network model. This hypothesis works with the assumption that regular behavior of nodes revolves around an average. However, some anomalies may occur. The 3-lambda model is stochastic and uses the three parameters associated with the average behavior of the nodes. The growth of the network based on this model assumes that one step of the growth is an interaction in which both new and existing nodes are participating. In the paper we present the results of the analysis of a co-authorship network and formulate a hypothesis and a model based on this hypothesis. Later in the paper, we examine the outputs from the network generator based on the 3-lambda model and show that generated networks have characteristics known from the environment of real-world networks.
\end{abstract}


	
\section{Introduction}
Network analysis is a phenomenon that affects research in many areas. One of the goals of network analysis is to describe the phenomena, properties, and principles that are universal and manifest in nature, society, and in the use of technology.
As a network, we understand an ordered pair $G = (V, E)$ (undirected unweighted graph) of a set $V$ of nodes and a set $E$ of edges which are unordered pairs of nodes from $G$.
The complex network structure can be viewed from the perspective of the network's state at the time of the analysis. Networks can, therefore, be described by the properties known from the environment of real-world networks, including, in particular, the small-world, free-scale, high average clustering coefficient, assortativity \cite{newman2002assortative}, community structure, shrinking diameter \cite{leskovec2005graphs}, but also others such as core-periphery structure \cite{rombach2014core} and self-similarity \cite{song2005self}. Underlying processes that take place during the evolution of real-world networks are also examined. Some models are based on analyzing these processes, which allows using the formally described underlying process as a generative mechanism. Such a mechanism can generate networks possessing one or more known properties. Models that reveal key principles include those using the preferential attachment to generate network centers \cite{albert2002statistical} or triadic closure i.e. completing interconnections into triangles, capable of generating community structure \cite{bianconi2014triadic}.

Models that provide key knowledge about networks are usually inherently simple. Most of them, however, focus on the question \textit{``How to connect a new node into the network?''} Our question is, \textit{``How does an existing node behave to its neighbors and other existing and new nodes during the network's evolution?''}. The result of our focus on the behavior of \textit{existing nodes} is a simple model \textit{without memory} and with only \textit{three parameters}. This new 3-lambda model is inspired by the evolution of the co-authorship network. For the analysis we used a network generated from a DBLP dataset and we worked with the assumption that in each publication is just one key author who picks out additional co-authors. In the analytically oriented experiment, we show that with such an assumption, the number of publications with a given number of co-authors corresponds approximately to a Poisson distribution. Based on the result of this experiment, we formulate a simple hypothesis and the resulting network growth model. This hypothesis assumes that one-step of the network growth is an interaction involving existing and new network nodes. In this respect our approach is similar to the model of collaborative networks published by Ramasco et al. \cite{ramasco2004self} and inspired by the analysis of co-authorship ego networks in research of Arnaboldi et al. \cite{Arnaboldi2016analysis}. 3-lambda is a stochastic model that estimates the number of nodes in the interaction using the Poisson distribution. In the experimental part of this paper, we describe the network generator based on our model and three experiments. The first experiment shows that generated networks have characteristics known from real-world networks, and how the selected properties change with a different setting of the generator. The second experiment shows how the properties of generated network change during its growth. The aim of the third experiment is to compare some characteristics of the DBLP network and large-scale generated networks.

The rest of the paper is organized as follows: Section \ref{sec:rel} focuses on the related work. Section \ref{sec:dblp} provides our findings and hypothesis on the real-world network extracted from the DBLP dataset. In Section \ref{sec:met}, we describe the 3-lambda model of collaborative network and network generator based on this model. Section \ref{sec:exp} focuses on three experiments with generated networks. We conclude and briefly discuss open problems in Section \ref{sec:conc}.

\section{Related work}
\label{sec:rel}
 
In the last two decades, the analysis of real-world networks has received extraordinary attention. One of the sources of data is social networks, which are growing at an enormous rate. Notable and a long investigated source in this area are co-authorship and, in general, collaborative networks. A common feature of this type of network is that underlying processes proceed in cliques, which then become a fundamental building block of the network. 
Barabasi et al. \cite{ barabasi2002evolution} presented and analyzed in detail a network model inspired by the evolution of co-authorship networks. The research presented by Ramasco et al.\cite{ ramasco2004self} falls into the same area; it analyzed in detail the development of collaboration networks. Their model combines preferential edge attachment with the bipartite structure and depends on the act of collaboration. The rise of the giant connected component in the set of $k$-cliques of a classical random graph was described by Derenyi at al. \cite{derenyi2005clique} as well as a $k$-clique community, as a union of all $k$-cliques. A novel model of multi-layer network was proposed by Battiston et al. \cite{battiston2016emergence}, their model captures a multi-faceted character of actors in collaborative networks. 

The universally recognized principle is the so-called preferential attachment. At the moment of the connecting of new nodes to the network during its growth, there is a preference for selecting high degree nodes. 
The well-known Barabasi-Albert model \cite{albert2002statistical} is based on experimental work and analysis of large-scale data. 
Zuev et al. \cite{zuev2015emergence} described how preferential attachment together with latent network geometry explains the emergence of soft community structure in networks and non-uniform distribution of nodes.

One of the basic characteristics of some types of networks (social and biological), is their community structure. Understanding the principles upon which communities emerge is a key task.
Growing network model using the triadic closure mechanism is able to display a nontrivial community structure, as was proposed by Bianconi at al. \cite{bianconi2014triadic}. 
The addition of links between existing nodes having a common neighbor as a local process leads to the emergence of preferential attachment as is stated by Shekatkar \& Ambika \cite{shekatkar2015complex}.
In another model for growing network proposed by Toivonen et al. \cite{toivonen2006model}, communities arise from a mixture of random attachment and implicit preferential attachment. 

A frequent feature of these approaches is that communities rise from a combination of links between existing nodes with their neighbors to new nodes. Some of these approaches do not use the preferential attachment for node selection because the scale-free property is the result of underlying processes.

Another well-known property of real-world networks is that communities have overlaps.
A node may belong to more cliques simultaneously, and this property is the basis of the Clique Percolation Method presented by Palla et al. \cite{palla2005uncovering}.
The clique graph, wherein cliques of a given order are represented as nodes in a weighted graph, is a conceptual tool to understand the $k$-clique percolation described by Evans \cite{evans2010clique}.
Yang and Leskovec introduced the Community-Affiliation Graph \cite{yang2012community} based on observation, that community overlaps are denser than communities themselves. 

Processes in the networks take place in time. Networks are from this perspective temporal, and each interaction is reflected in changes to the network structure. Application of the principles mentioned above in the course of network evolution allows us to examine how network structure changes over time. 
Holme \& Saramaki \cite{holme2012temporal} presented a time-varying importance of nodes and edges together with a survey of existing approaches and the unification of terminology in the area of temporal networks research.
Ramasco at al. \cite{ramasco2004self} studied social collaboration networks as dynamic networks growing in time by the continuous addition of new acts of collaboration and new actors.
In real-world networks, particularly social ones, instances often have strong relations defined as interactions that are frequently repeated (nodes remember them), as well as weak relations representing the occasional interactions. Karsai et al. \cite{karsai2014time} explain, how is creating new relationships and strengthening existing links in networks important for network evolution. 

There are also novel approaches focused on models which allow generating networks with predictable properties. For instance, Zhang et al. \cite{zheng2014simple} formulated a generative model as an optimization problem.

In our approach, we do not use preferential attachment in a straightforward manner. It is, however, a side effect of principles related to the nature of the formation of the community structure. In our model, we use a clique as a structural element of the network. A clique is the result of interaction among nodes and is the basis of the community structure. Our networks are generated as temporal because one interaction is the result of one step of the growth of the network.

\section{DBLP Dataset Analysis} 
\label{sec:dblp}

We studied the DBLP dataset which contains basic bibliographic information of publications from the computer science field. This data is freely available\footnote{\url{http://www.informatik.uni-trier.de/~ley/db/}} and contains highly relevant information about publication activity from the period of nearly fifty years, even though they are not complete. At the time we downloaded this dataset (July 2016), and after first pre-processing, it contained a total of $2988015$ publications. Further characteristics can be seen in Table \ref{tab:dblp}. 
\begin{table}[htbp]
  \centering
  \caption{DBLP dataset}
    \begin{tabular}{|l|r|} 
		\hline
    Total number of publications & 2988015 \\
    Total number of authors & 1622828 \\
    Mean papers per author & 5.113 \\
    Mean authors per paper & 2.895 \\
    \hline
    \end{tabular}%
  \label{tab:dblp}%
\end{table}%

Based on the co-authorship of authors, we constructed a social network where authors are linked if they co-authored a paper. The weight of the edge corresponds to the number of co-authored papers. Basic characteristics of this network and its maximal connected component are in Table \ref{tab:dblpcompo}.

\begin{table}[htbp]
  \centering
  \caption{DBLP network and its max. connected component}
		\begin{tabular}{|l|rr|}
    \hline
          & $net$ & $max CC$\\
    \hline
    Total number of nodes & 1554772 & 1408331 \\
    Total number of edges & 6930545 & 6768108 \\
    Density & 5.73\mbox{\sc{e}-}06 & 6.83\mbox{\sc{e}-}06 \\
    Mean degree & 8.915 & 9.612 \\
    Number of connected components & 49459 & 1 \\
    Global clustering coefficient & 0.1749 & 0.1741 \\
    Mean local cluster coefficient & 0.7341 & 0.7217 \\
    Mean edge weight & 1.746 & 1.762 \\
    Number of communities (Louvain) &   -    & 422 \\
    \hline
    \end{tabular}%
  \label{tab:dblpcompo}%
\end{table}%

In additional pre-processing of the dataset, we set each publication a month which corresponded to the date of a conference or date of publication in a journal, respectively.  If the record in the DBPL did not contain the month of publication, we chose the month for a given year randomly. Furthermore, we assumed that the author existing in a given month is the author who had at least one publication in the month preceding the given month. In the next step, we dropped all publications that did not contain any already existing authors. For the rest of publications, we have identified the main author, who was the first existing author in the order of the co-authors of these publications.

\begin{figure}[ht]
\centering
\includegraphics[width=\linewidth]{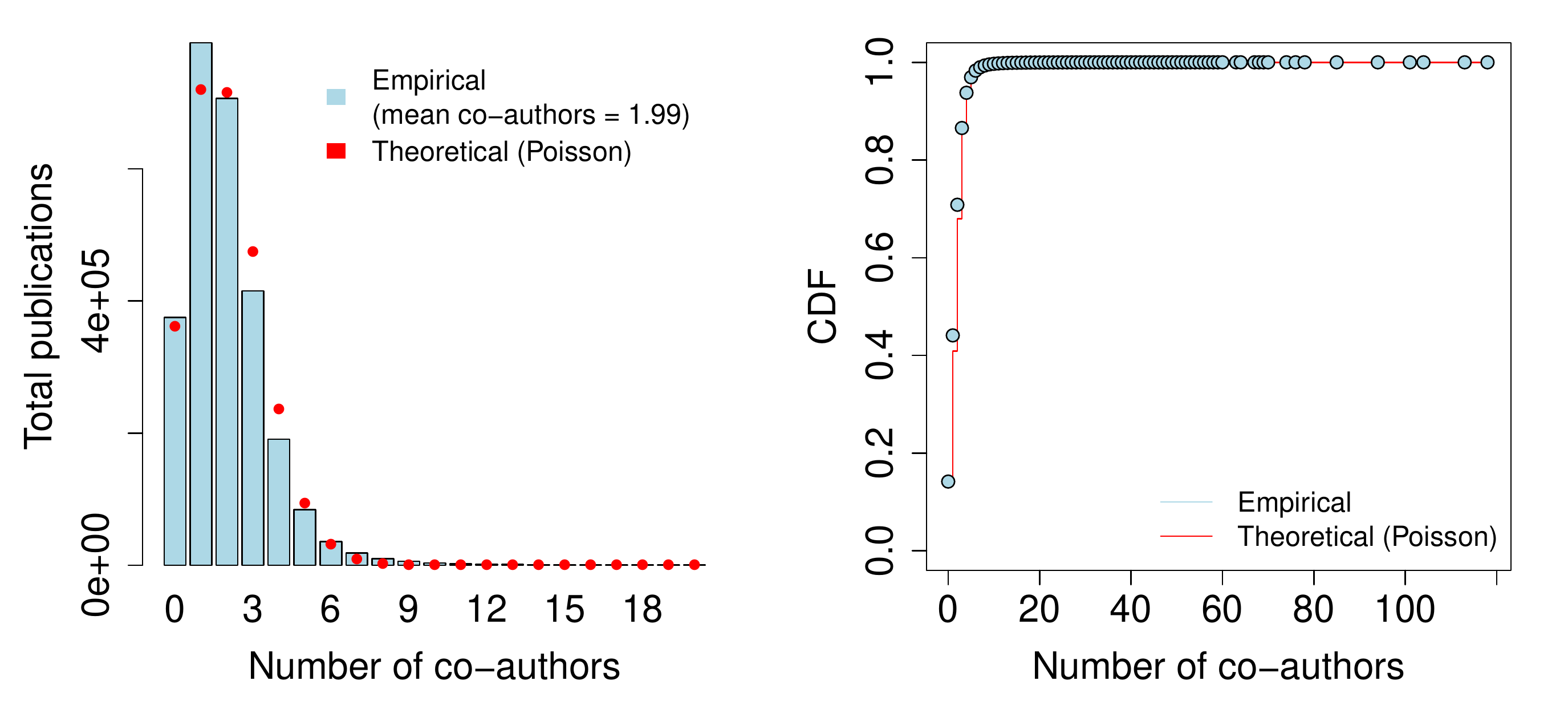}
  \caption{DBLP: Poisson distribution}\label{fig:poisfit}
\end{figure}

The first objective of the experiment with the DBLP dataset was to discover what shape has the distribution of the number of publications depending on the number of co-authors of the main author. Figure \ref{fig:poisfit} shows the distribution for the first twenty values (i.e. up to $20$ co-authors) and cumulative distribution of all values (except for one publication with $286$ authors). This distribution is compared to a Poisson distribution with a $\lambda$ value equal to the average number of co-authors, which is $1.99$.

\begin{figure}[ht]
\centering
\includegraphics[width=\linewidth]{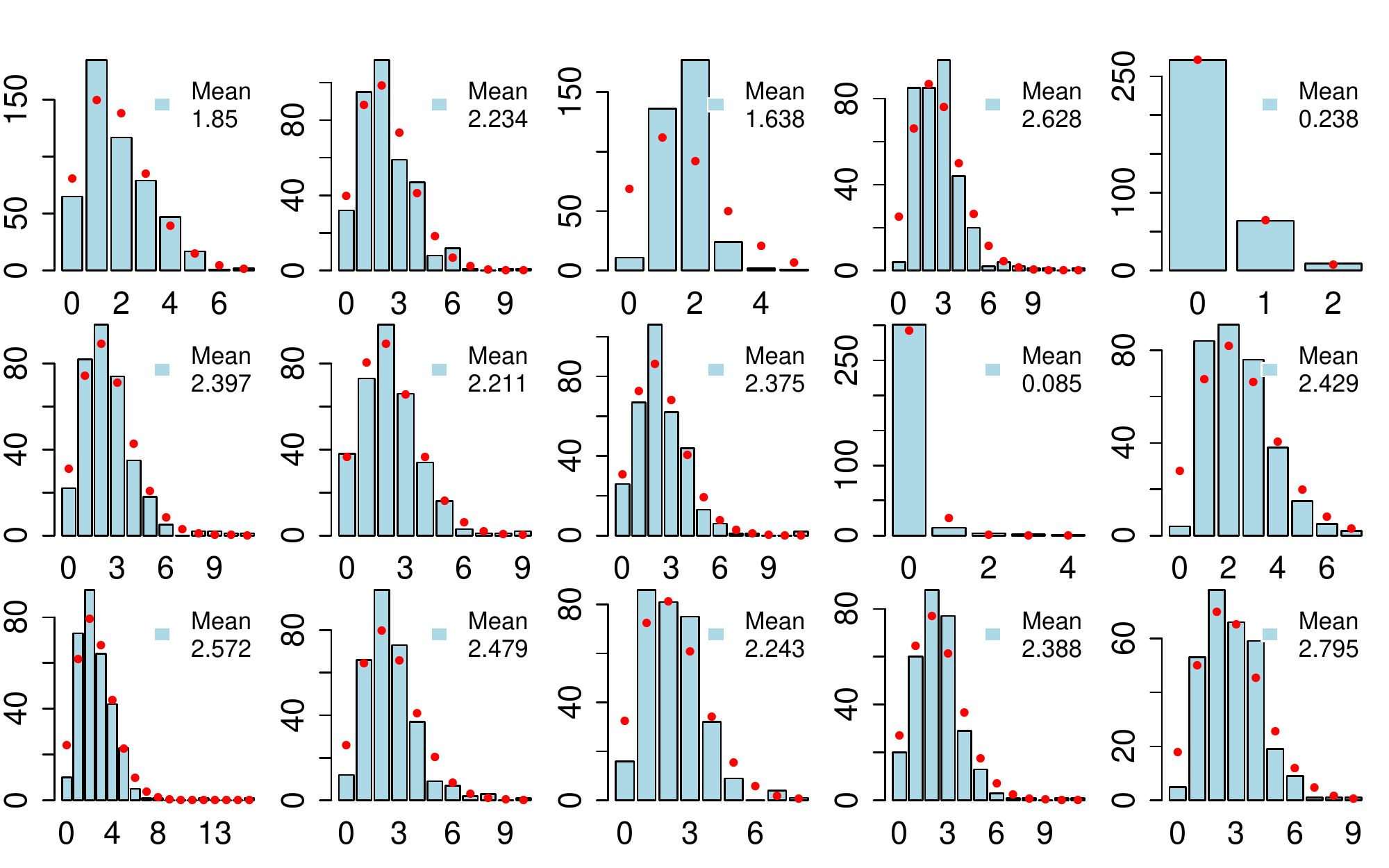}
  \caption{Co-authors histograms for top 15 authors (by number of publications)}
	\label{fig:top15}
\end{figure}

The second objective of the experiment was to find authors who are most often in the role of the main author of a publication. The results of this part of the experiment are to be understood only as an estimate based on the above-mentioned assumptions about the main author. Empirical distribution, together with the theoretical value of the Poisson distribution for the first fifteen authors with the highest number of publications in the role of the first author, is shown in Figure \ref{fig:top15}.

\subsection{Hypothesis}
Results of the analysis of the DBLP dataset can not be easily generalized. However, experiments have shown quite clearly that the probability of a publication with a certain number of co-authors follows the Poisson distribution. If we extend thoughts about the main author by what kind and how many different co-authors he/she has, we may divide co-authors into three groups. In the first group are authors with whom the main author has previously published. In the second group are those who have previously published but not yet together with the main author. In the third group are new authors, i.e. those who have no previous publications. We can then formulate a hypothesis, based on which we define the new network model in the next part of the paper.

\bigbreak
\textbf{Hypothesis}
\begin{enumerate}
	\item The variables describing the number of co-authors in different groups are independent random variables.
  \item Just as the total number of co-authors, these variables follow Poisson distribution.
\end{enumerate}

In this paper, we are working with a co-authorship network, which is essentially a collaborative network. Therefore, the presented model can not be taken as a universal model. The model is primarily about a simple simulation of the development of collaborative networks inspired by analyzing the co-authorship network.

\section{3-lambda Model} 
\label{sec:met}

\begin{figure*}[ht]
\centering
 \begin{subfigure}{17em}
  \centering
	\includegraphics[width = 16em]{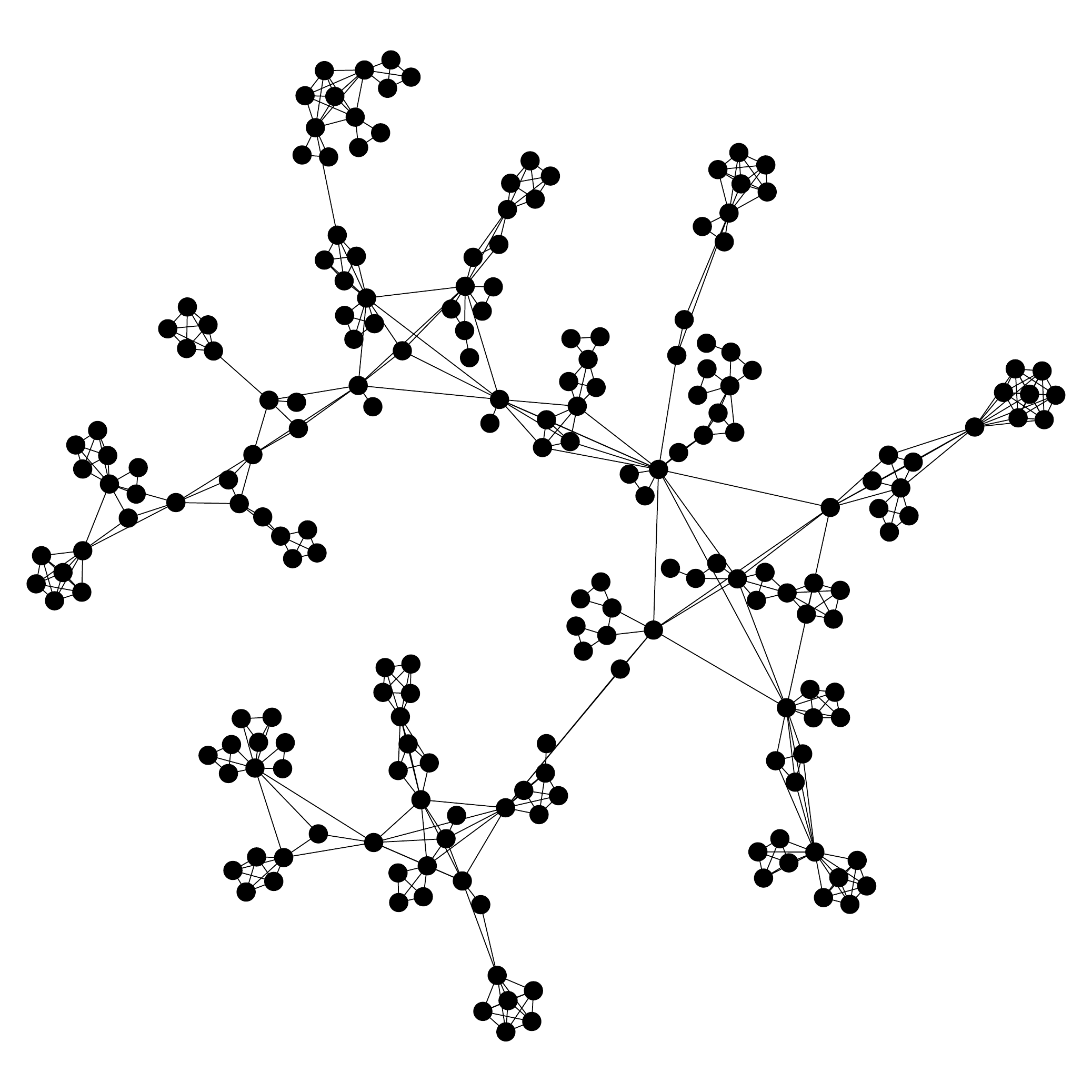}
\caption{$\lambda_1$ = 0, $\lambda_2$ = 3, $\lambda_3$ = 0}
\label{fig:lambda2net}
 \end{subfigure}
 \begin{subfigure}{17em}
	\centering
		\includegraphics[width =16em]{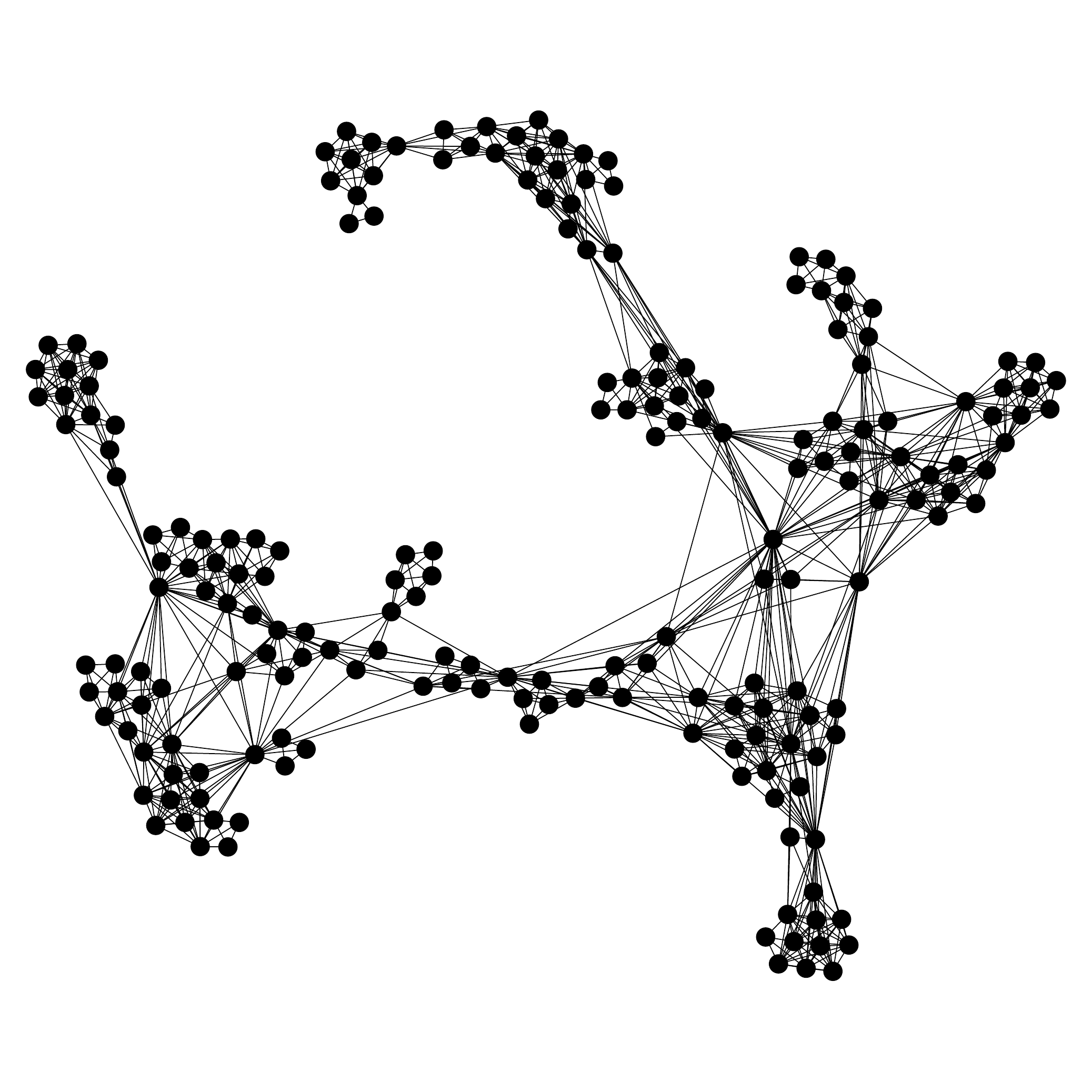}
		\caption{$\lambda_1$ = 2, $\lambda_2$ = 3, $\lambda_3$ = 0}
		\label{fig:lambda1net}
  \end{subfigure}
  \begin{subfigure}[ht]{17em}
		\centering
			\includegraphics[width =16em]{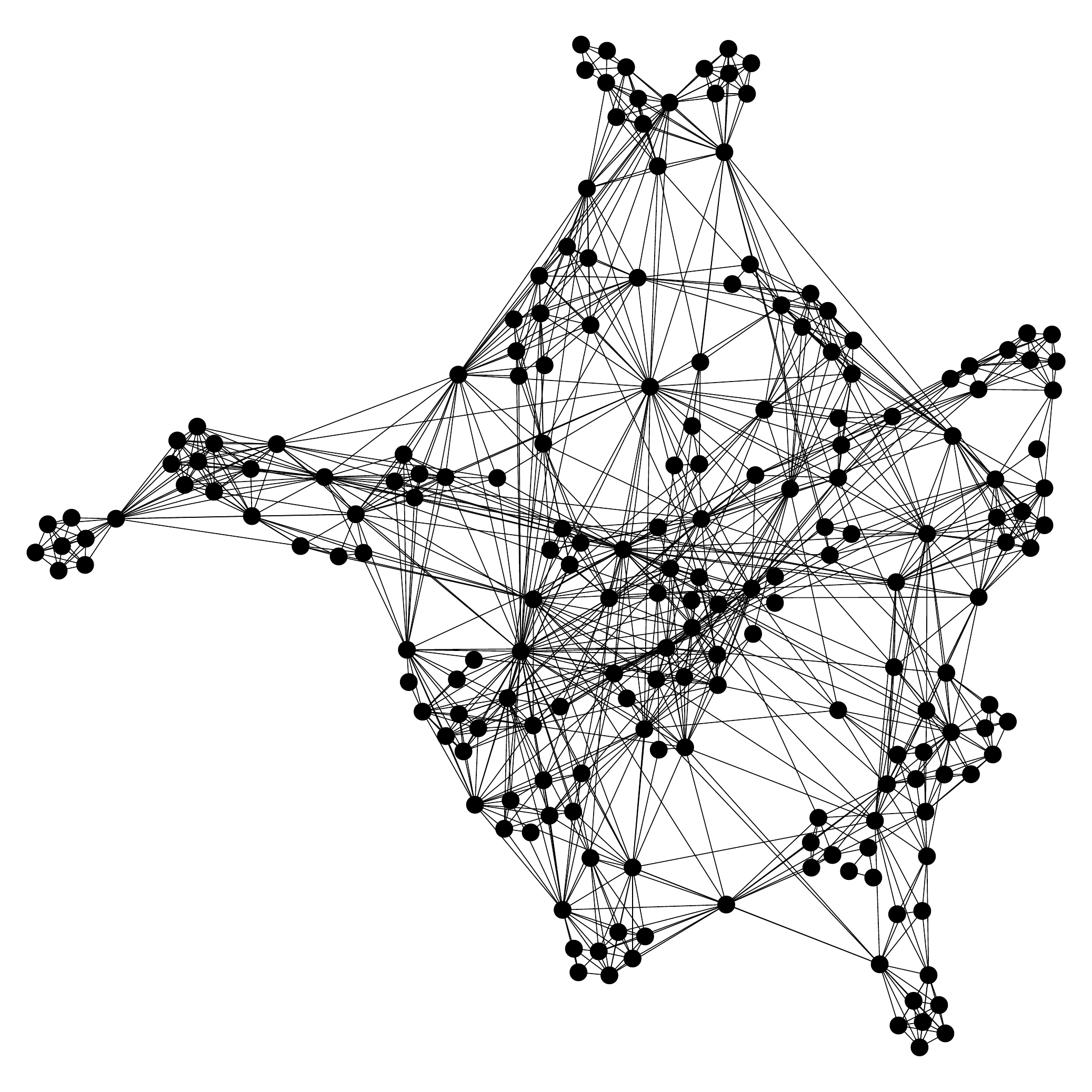}
			\caption{$\lambda_1$ = 2, $\lambda_2$ = 3, $\lambda_3$ = 0.5}
			\label{fig:lambda3net}
  \end{subfigure}
	\caption{3-lambda model: generated networks with $200$ nodes}
\label{fig:L3D}
\end{figure*}

As mentioned in the introduction, the model is based on the assumption that one step of the network's growth is one interaction. Both new and existing nodes are involved in the interaction, and after the interaction, there are edges between all involved pairs of nodes. Some (or all) of these edges between pairs of involved nodes could exist before the interaction. The model is inherently temporal since we can associate with each node or edge the list of interactions in which they participated during the growth of the network (the list of interactions is a list of steps of the growth in the order in which the interactions took place). It is further assumed that neither nodes nor edges age, so they are not removed during the growth of the network. From this perspective the model is growing.

In the model, nodes involved in the interaction have four different roles:

\begin{enumerate}
	\item key node of the interaction (proactive)
  \item nodes adjacent to the proactive node (neighbors)
  \item new nodes (newbies)
  \item nodes that are not adjacent to the proactive node (new connections)
\end{enumerate}

It is essential for the model that each interaction always contains exactly one proactive node and may (but doesn't have to) include nodes in three other roles. How many spots will be represented by each of these three roles is selected based on the Poisson distribution with a preselected $\lambda_1$ (neighbors); $\lambda_2$ (newbies); and $\lambda_3$ (new connections). Furthermore, the model assumes that the selection of specific existing nodes for neighbor roles (in the number equal to the maximum number of all neighbors at most) and new connections is random.


A natural characteristic of the model is that nodes with a higher degree are more likely to participate in an interaction in which the proactive node has at least one neighbor. Although a proactive node of the interaction and its neighbors are being selected at random, a high degree node has a greater chance of being selected as a neighbor of the proactive node.

$\lambda_1$, $\lambda_2$ and $\lambda_3$ significantly affect the density of the network. If we assume that a randomly selected interaction has, based on the corresponding distributions, $b$ neighbors of a proactive node, $n$ new nodes and $e$ nodes unconnected to the proactive node, then the number of nodes involved in this interaction (interaction size) is in Equation \ref{eq:xdef} 

\begin{equation} s = 1 + b + n + e \label{eq:xdef}
\end{equation}

and the following applies:

\begin{itemize}
	\item $n$ new nodes, which must connect to a proactive node and each other, is created.
	\item There are $b$ nodes adjacent to the proactive node which must first get connected among each other, then with $e$ nodes that are not adjacent to the proactive node (these edges may already exist prior to the interaction). Finally, they must connect with the new $n$ nodes.
	\item There are $e$ nodes not adjacent to the proactive node, which must get connected with it. Next, they must get connected among each other (these edges may already exist prior to the interaction) and $n$ new nodes.
\end{itemize}


The minimum and the maximum number of new connections (edges) is in Equations \ref{eq:min} and \ref{eq:max}, respectively.

\begin{equation}	MIN = n + e + b \cdot n + e \cdot n + \frac{n \cdot (n - 1)}{2} \label{eq:min}
\end{equation}

\begin{equation} MAX = MIN + e \cdot b + \frac{b \cdot (b - 1)}{2} + \frac{e \cdot (e - 1)}{2} \label{eq:max}
\end{equation}

Thus, if for example we set $b = 2, n = 3, e = 1$ for the interaction, then:

\begin{itemize}
	\item interaction size is $s = 1 + 2 + 3 + 1 = 7$
	\item after the interaction a total of $21$ edges exist between pairs of nodes
	\item $MIN = 3 + 1 + 6 + 3 + 3 = 16$ in the case of $5$ edges existing between pairs of nodes prior to the interaction
	\item $MAX = 10 + 2 + 1 + 0 = 19$ in the case of $2$ edges existing between pairs of nodes prior to the interaction
\end{itemize}

Each of $\lambda_1$, $\lambda_2$ and $\lambda_3$ affects different property of the network generated by 3-lambda model.

%
%
%
%
%

\begin{itemize}
	\item $\lambda_2$ (newbies) defines the growth rate of the network and provides a tree-like network structure, see Fig. \ref{fig:lambda2net}. To construct a network with $N$ nodes requires approximately $N/$$\lambda_2$ interactions.
	\item $\lambda_1$ (neighbors) constitutes network community structure through local connections of existing neighbors and new nodes, see Fig. \ref{fig:lambda1net}.
	\item $\lambda_3$ (new connections) ensures linking of nodes that are not adjacent, thereby linking communities. The consequence is emerging of core-periphery network structure, see Fig. \ref{fig:lambda3net}.
\end{itemize}

%

\subsection{Network Generator}
\label{sec:alg}
The network generator uses a simple algorithm which comes directly from the model description. The only extra step is setting up the initial network state. The model is memory-less, which allows working with an arbitrary initial state. For our generator we chose a complete graph with a number of nodes equal to the round of $(1 + \lambda_1 + \lambda_2 + \lambda_3)$ as the default state. Algorithm \ref{alg:1} describes the whole process of generating a network. The generated network is a connected graph; the algorithm starts with a complete graph and each interaction contains at least one existing node. In our implementation, for reasons of analysis and visualization, we store the number of interactions for each node and edge, which is not mentioned in the algorithm.

\begin{algorithm}
\SetKwInOut{Input}{input}\SetKwInOut{Output}{output}
\Input{number of nodes $N$, $\lambda_1, \lambda_2, \lambda_3$}
\Output{generated network $G$}
\SetKwIF{If}{ElseIf}{Else}{if}{then}{else if}{else}{endif}
\SetAlgoLined
choose $s = ROUND(1 + \lambda_1 + \lambda_2 + \lambda_3)$\\
create $G = (V, E)$  as a complete graph with $s$ nodes\\

\While{$V$ has less than $N$ nodes}
{ 
  choose from $V$ randomly proactive node $A$ \\
  $b =$ number of neighbors by Poisson($\lambda_1$) ($b$ is the number of neighbors of $A$ at most) \\
  $n =$ number of newbies by Poisson($\lambda_2$)\\
  $e =$ number of new connections by Poisson($\lambda_3$)\\

  create a list $I$ with a proactive node $A$\\
  add to list $I$ $b$ randomly selected neighbors of node $A$ from $V$\\
  create $n$ new nodes, add them to $V$ and $I$\\
  add to list $I$ $e$ randomly selected not-neighbors of node $A$ from $V$ (if such nodes exist)\\

 \ForEach{pair of nodes $(V_i,V_j) \in I$}
 {
    \If{no edge $e$ between $V_i$ and $V_j$ exists}
    {
      create $e$ and add to $E$\\
    }
 }
}

\caption{3-lambda model network generator}\label{alg:1}
\end{algorithm}

The average number of nodes in an interaction is approximately  $s = 1 + \lambda_1 + \lambda_2 + \lambda_3$. However, the average is slightly lower because the number of neighbors selected for interaction through the simulation of the Poisson distribution is limited by the actual (maximum) number of neighbors of the proactive node.

The complexity of the algorithm is $O (s^2 \cdot \frac{N}{\lambda_2})$, which is based on the fact that:

\begin{itemize}
	  \item To generate $N$ nodes requires approximately $N / \lambda_2$ interactions ($\lambda_2$ is the average number of new nodes in one interaction).
    \item The number of edges between nodes in an interaction is in quadratic relation to the number of nodes of this interaction (the interaction takes place in a complete sub-graph with $n$ nodes and $m = \frac{n * (n-1)}{2}$ edges).
\end{itemize}

The calculation of algorithm complexity does not include the complexity of the simulation of the Poisson distribution for individual \textit{lambdas} (in our case, the Knuth's algorithm with complexity $O(\lambda)$ was used).

\section{Experimental Evaluation}
\label{sec:exp}
We are using using three different settings, $Setting = [\lambda_1, \lambda_2, \lambda_3]$, in the experiments. In $Setting_1 = [1.6, 0.35, 0.05]$, the average interaction involved three nodes (a triad), which corresponds approximately to the analyzed co-authorship network. This setting presumes the interaction is dominated by neighbors of the proactive node with the occasional participation of new nodes and rather exceptional participation of existing nodes yet not connected to the proactive node. In $Setting_2 = [3, 6, 1]$ predominate new nodes, and the average number of nodes in interaction is $11$. In $Setting_3 = [0.45, 0.45, 0.1]$ interaction involved two nodes (a dyad) on average, wherein the number of neighbors and new nodes are balanced and new connection occurrences are less likely.

\begin{figure}[ht]
	\centering
  \begin{subfigure}{2.7cm}
    \centering\includegraphics[width=2.5cm]{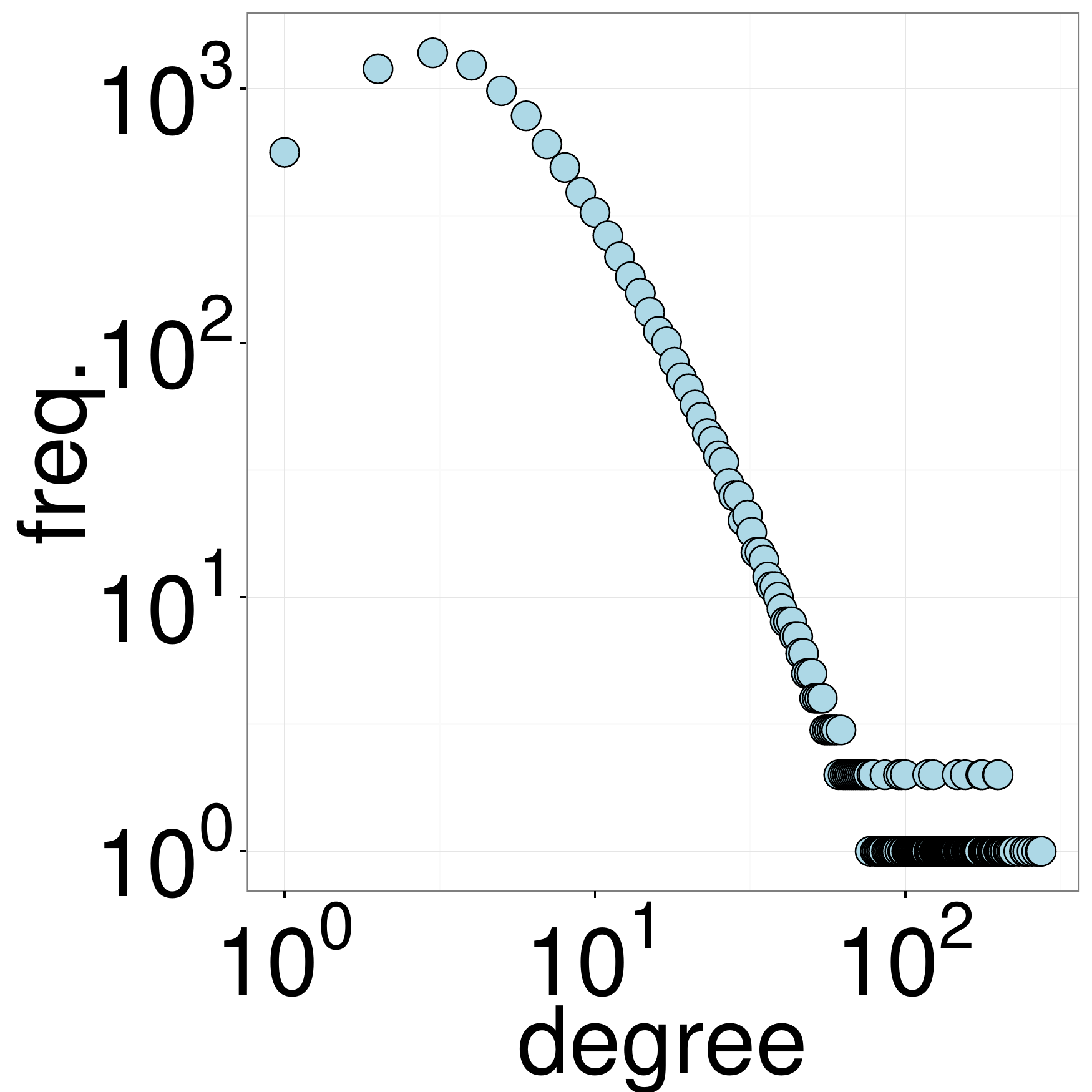}
    \caption{$Setting_1$}
  \end{subfigure}
  \begin{subfigure}{2.7cm}
    \centering\includegraphics[width=2.5cm]{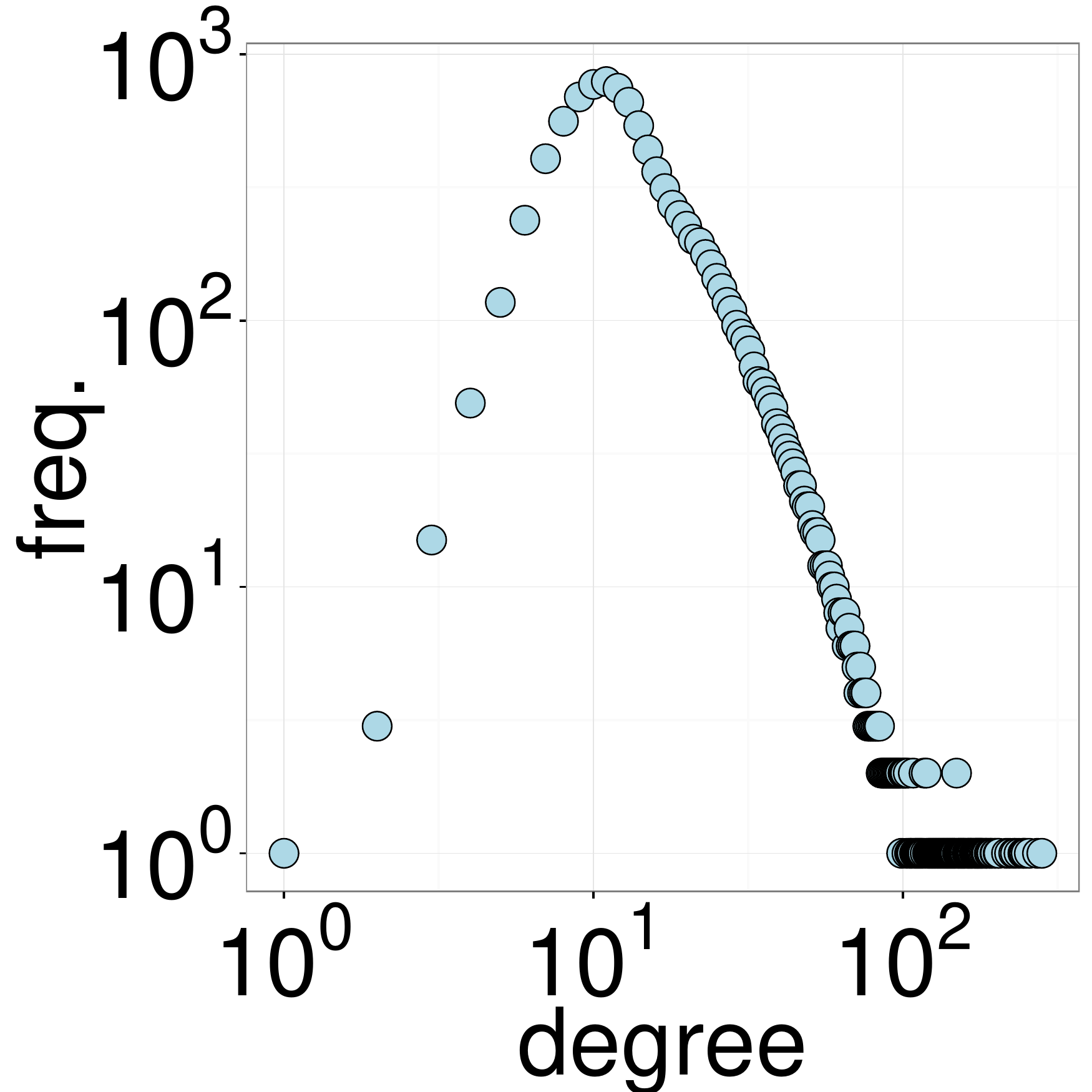}
    \caption{$Setting_2$}
		  \end{subfigure}
   \begin{subfigure}{2.7cm}
    \centering\includegraphics[width=2.5cm]{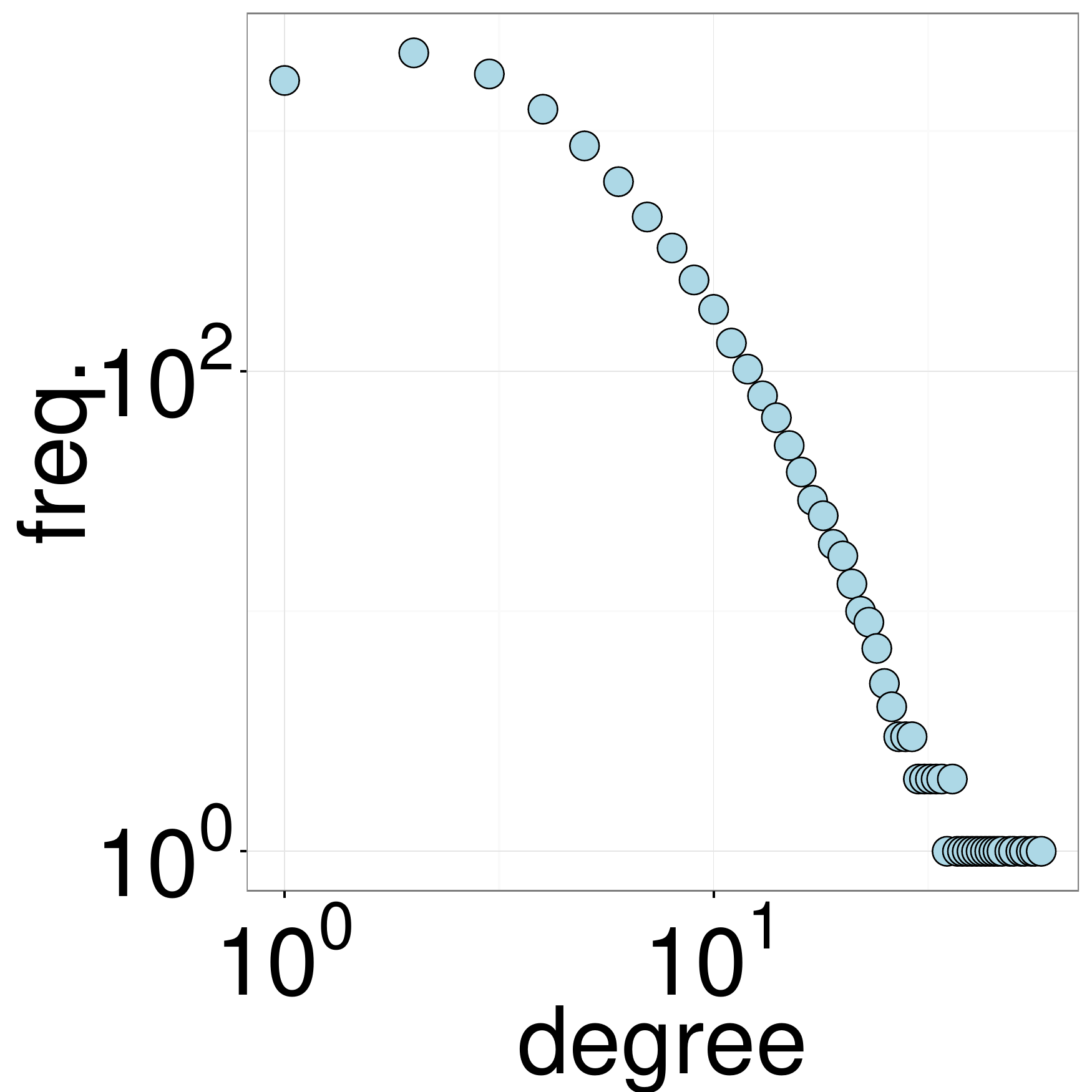}
    \caption{$Setting_3$}
  \end{subfigure}
	\caption{Degree distribution}
\label{fig:DD}
\end{figure}

\begin{figure}[ht]
	\centering
  \begin{subfigure}{2.7cm}
    \centering\includegraphics[width=2.5cm]{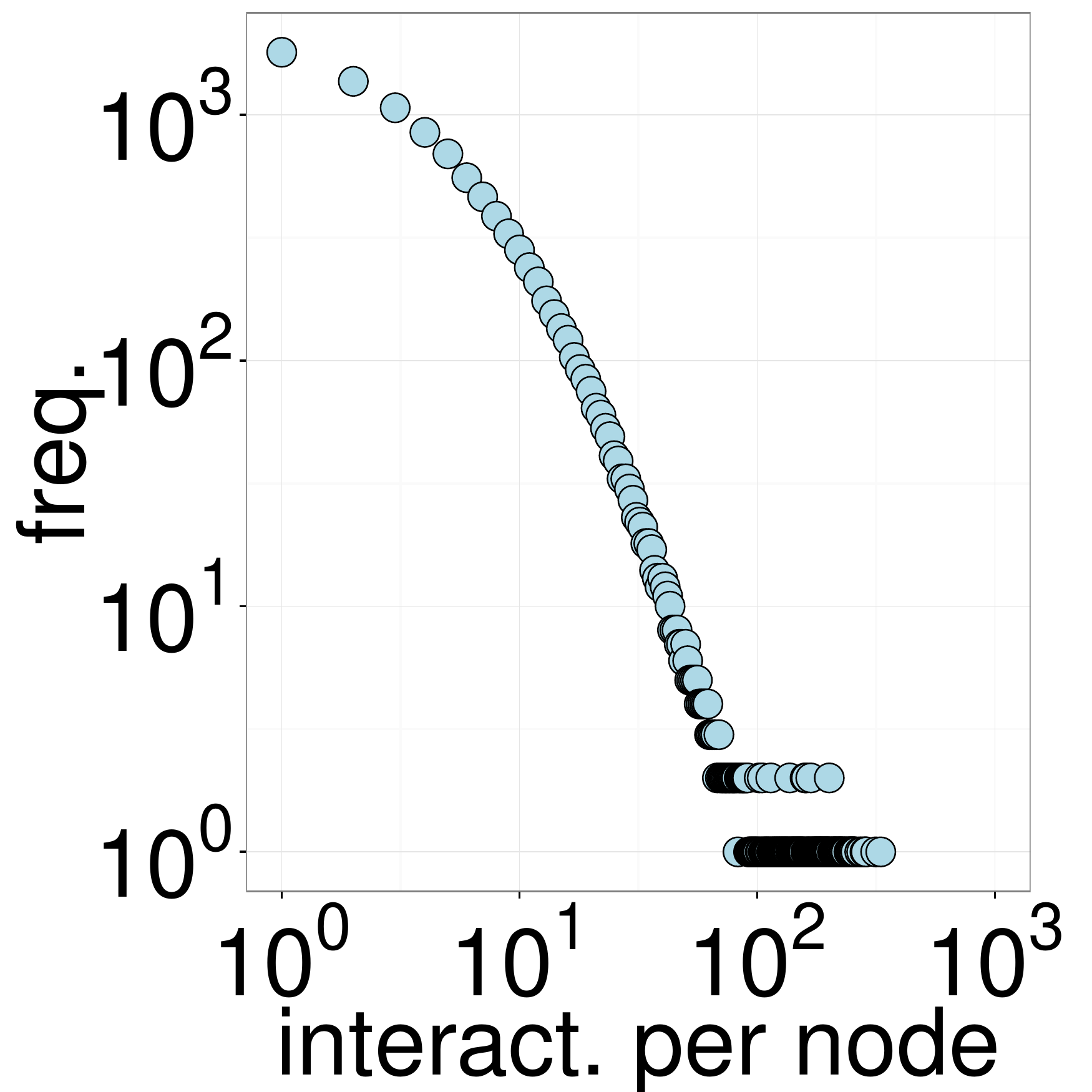}
    \caption{$Setting_1$}
  \end{subfigure}
  \begin{subfigure}{2.7cm}
    \centering\includegraphics[width=2.5cm]{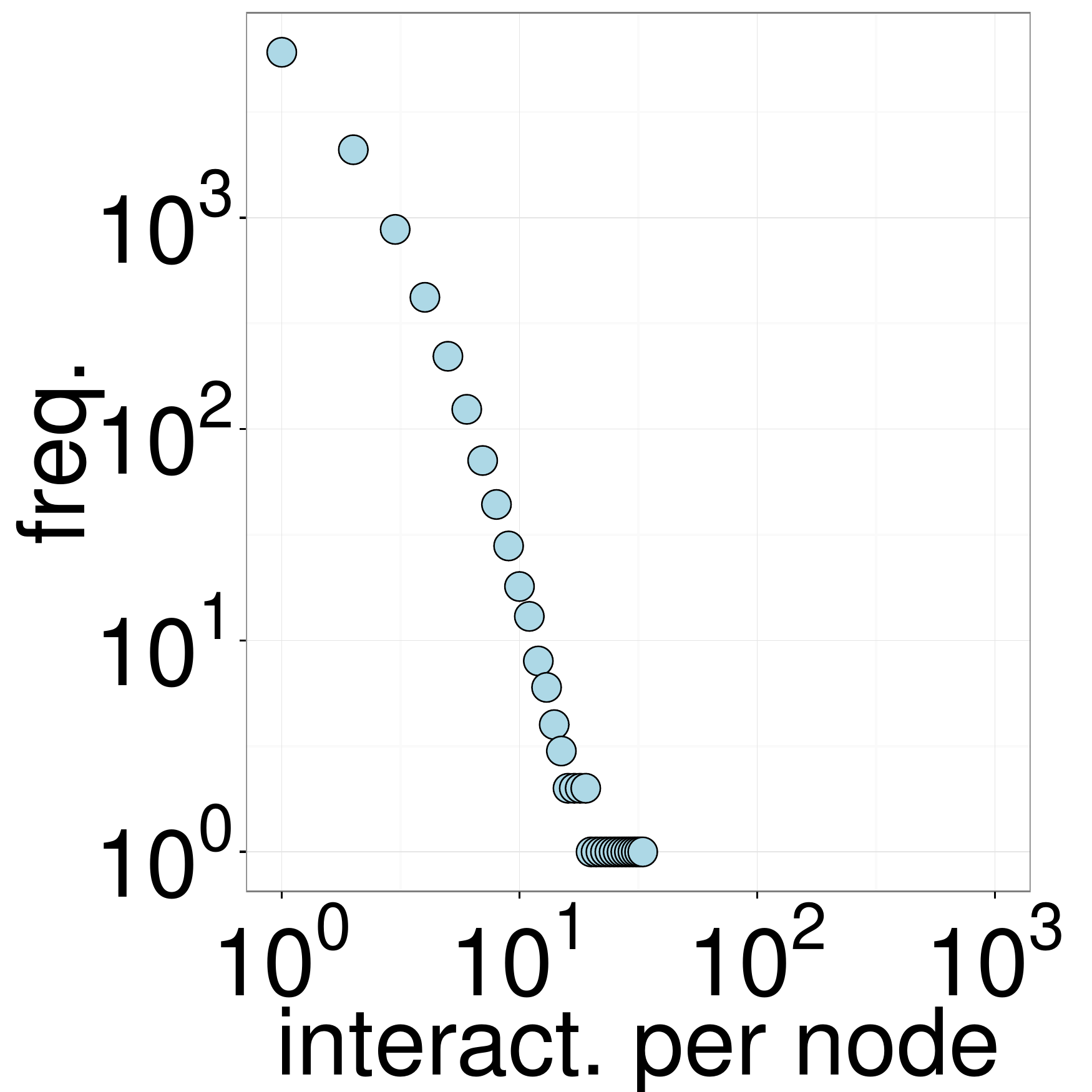}
    \caption{$Setting_2$}
		  \end{subfigure}
   \begin{subfigure}{2.7cm}
    \centering\includegraphics[width=2.5cm]{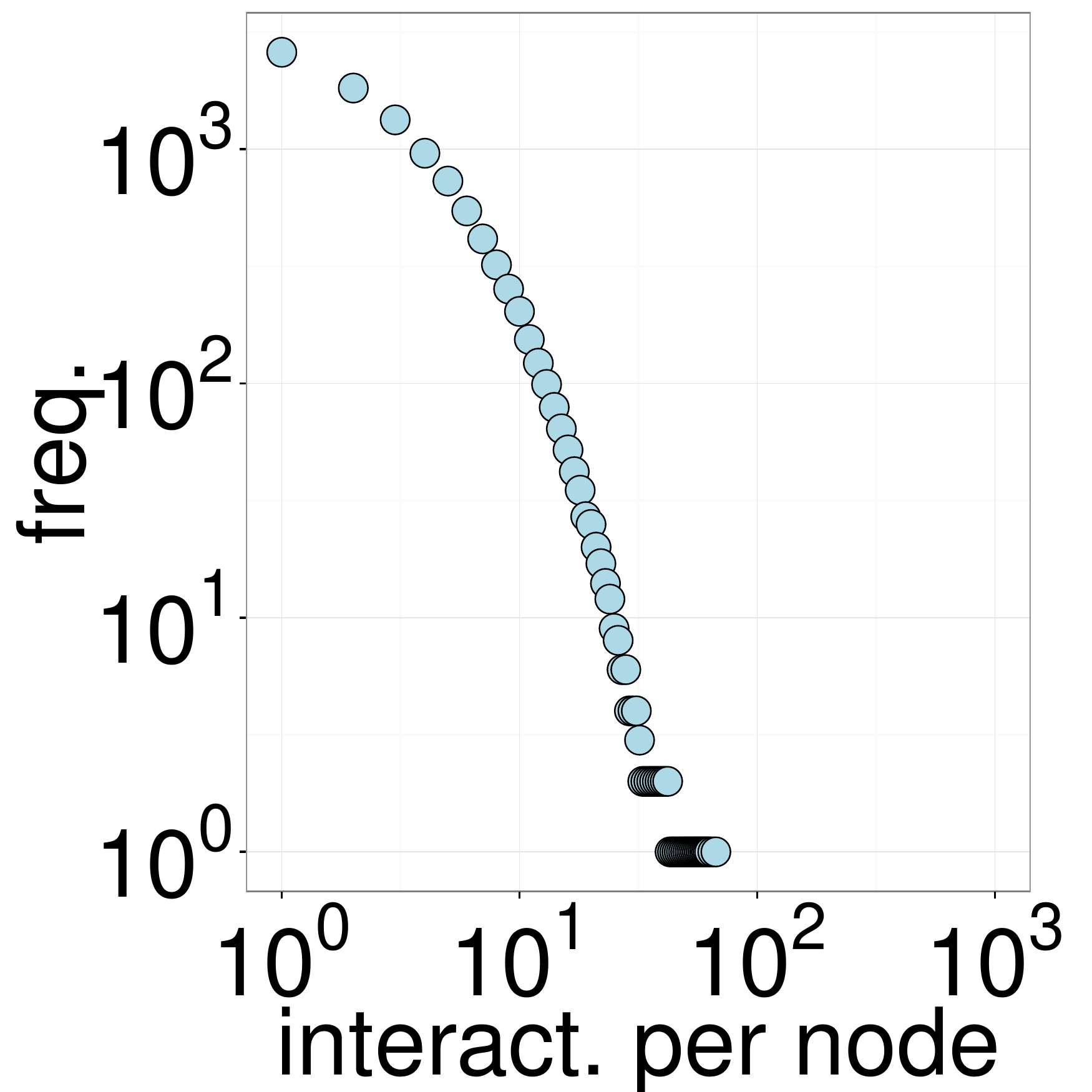}
    \caption{$Setting_3$}
  \end{subfigure}
	\caption{Interactions per	node}
\label{fig:IpN}
\end{figure}

\begin{figure}[ht]
	\centering
  \begin{subfigure}{2.7cm}
    \centering\includegraphics[width=2.5cm]{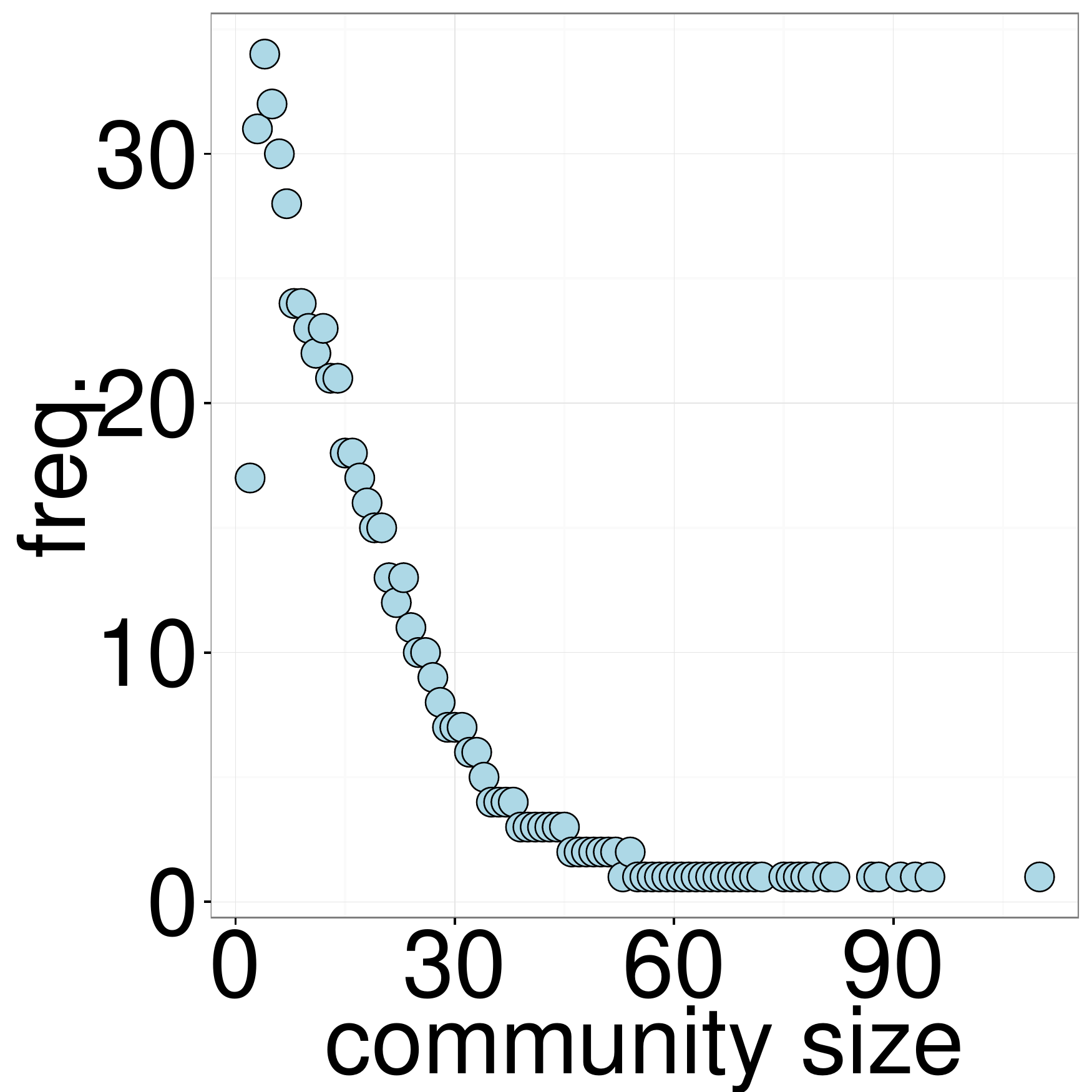}
    \caption{$Setting_1$}
  \end{subfigure}
  \begin{subfigure}{2.7cm}
    \centering\includegraphics[width=2.5cm]{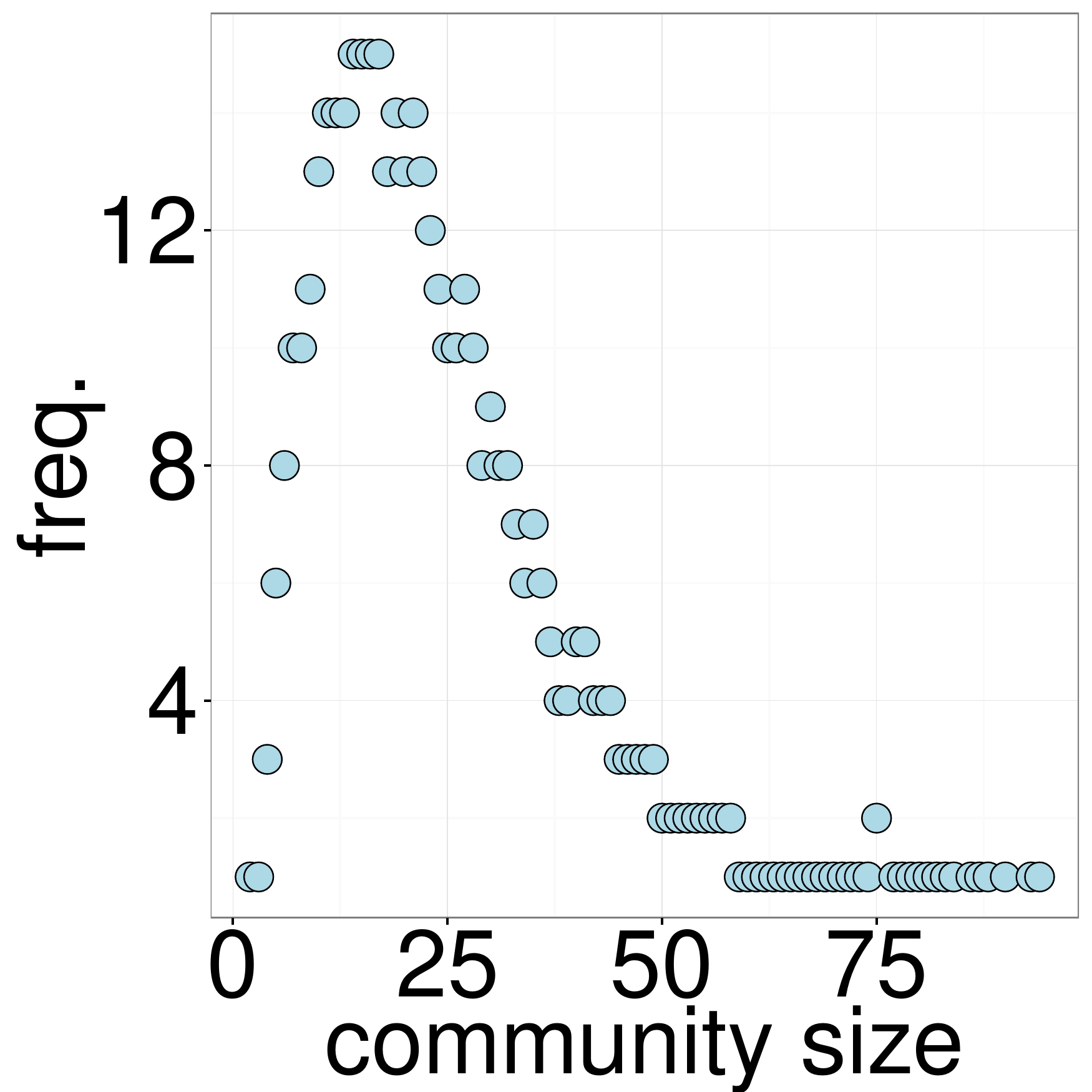}
    \caption{$Setting_2$}
		  \end{subfigure}
   \begin{subfigure}{2.7cm}
    \centering\includegraphics[width=2.5cm]{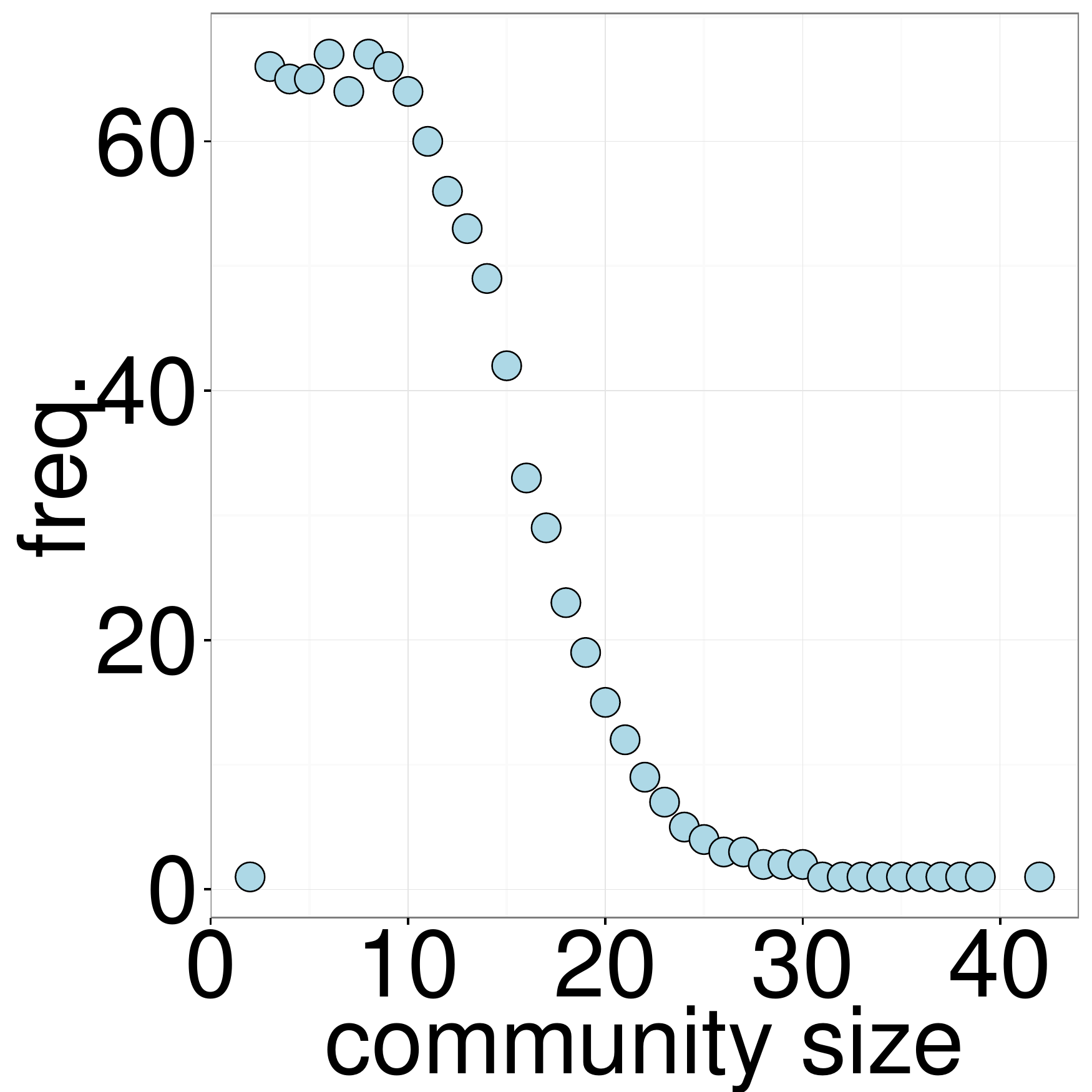}
    \caption{$Setting_3$}
  \end{subfigure}
	\caption{Community size}
\label{fig:ComSize}
\end{figure}

\subsection{Properties of Generated Network}

\begin{table*}[ht]
  \centering
  \caption{Global properties}
		\begin{tabular}{|c|r|rrrrrrrrrrr|}
\hline
Setting &  & $n$ & $m$ & $<$k$>$ &  $<$l$>$ & $l_{max}$ & $CC$ & $r$ & $com_{IM}$ & $com_L$ & $Q_{IM}$ & $Q_L$ \\ 
\hline
$Setting_1$ & mean & 10001.10 & 40108.97 & 8.0209 & 4.8101 & 12.07 & 0.6555 & 0.14522 & 609.40 & 54.70 & 0.6424 & 0.7119 \\ 
& sd & 0.31 & 398.80 & 0.0797 & 0.0421 & 0.64 & 0.0029 & 0.01769 & 12.37 & 7.04 & 0.0062 & 0.0085 \\ 
\hline	
$Setting_2$ & mean & 10003.87 & 88620.40 & 17.7172 & 4.2369 & 8.43  & 0.8087 & 0.12961 & 422.60 & 42.90 & 0.6772 & 0.7334 \\ 
& sd & 1.53 & 797.95 & 0.1600 & 0.0277 & 0.63  & 0.0018 & 0.00913 & 10.30 & 2.78 & 0.0049 & 0.0067 \\ 
\hline
$Setting_3$ & mean & 10001.17 & 21494.57 & 4.2984 & 6.8965 & 16.23 & 0.4784 & 0.19907 & 953.30 & 68.70 & 0.7276 & 0.8129 \\ 
&  sd & 0.46 & 142.64 & 0.0285 & 0.0609 & 0.63 & 0.0044 & 0.01222 & 15.55 & 3.19 & 0.0035 & 0.0038 \\ 
\hline
\end{tabular}%
  \label{tab:gp}%
\end{table*}%
\begin{table}[ht]
  \centering
  \caption{Interactions}
\begin{tabular}{|r|r|rrr|}
	\hline
Setting & & $I$ & $<$i$>$ & $<$s$>$ \\ 
\hline
$Setting_1$ & mean & 28561.90 & 8.0887 & 2.8323 \\ 
&  sd & 278.93 & 0.0817 & 0.0082 \\ 
\hline	
$Setting_2$ & mean & 1664.53 & 1.8278 & 10.9860 \\ 
& sd & 17.69 & 0.0115 & 0.0824 \\
\hline	
$Setting_3$ & mean & 22204.43 & 4.3772 & 1.9716 \\ 
& sd & 202.06 & 0.0292 & 0.0069 \\ 
\hline
	\end{tabular}%
  \label{tab:interact}%
\end{table}%

In the first experiment, we show the properties of networks generated with different settings. We generated for each setting $100$ networks with approximately $10000$ nodes. Table \ref{tab:gp} summarizes average values (and standard deviation) of measured properties for each setting. Measured properties include number of nodes $n$ and edges $m$, average degree $<$k$>$, average shortest path length $<$l$>$, diameter $L_{max}$, average clustering coefficient $CC$, assortativity $r$, number of communities detected by Infomap \cite{rosvall2009map} $com_{IM}$ and Louvain \cite{blondel2008fast} $com_L$ algorithm, and corresponding modularities $Q_{IM}$ and $Q_L$, respectively. Table \ref{tab:interact} contains average values associated with the temporality of the network, i.e. total number of interactions $I$, average number of interactions $<$i$>$ and average number of nodes in interaction $<$s$>$. Figures \ref{fig:DD}-\ref{fig:ComSize} show degree distribution, number of interactions distribution and distributions of community size detected by Infomap algorithm.

The experiment indicates that all three settings generate networks with small-world and scale-free characteristics. The first and second setting have generated networks of high average clustering coefficient. Networks have a tendency to be assortative. Assortativity values correspond to all settings to the values known from social networks \cite{newman2002assortative}. Generated networks also have community structure and a high modularity for all settings.

\begin{figure}[ht]
\centering
\includegraphics[width=\linewidth]{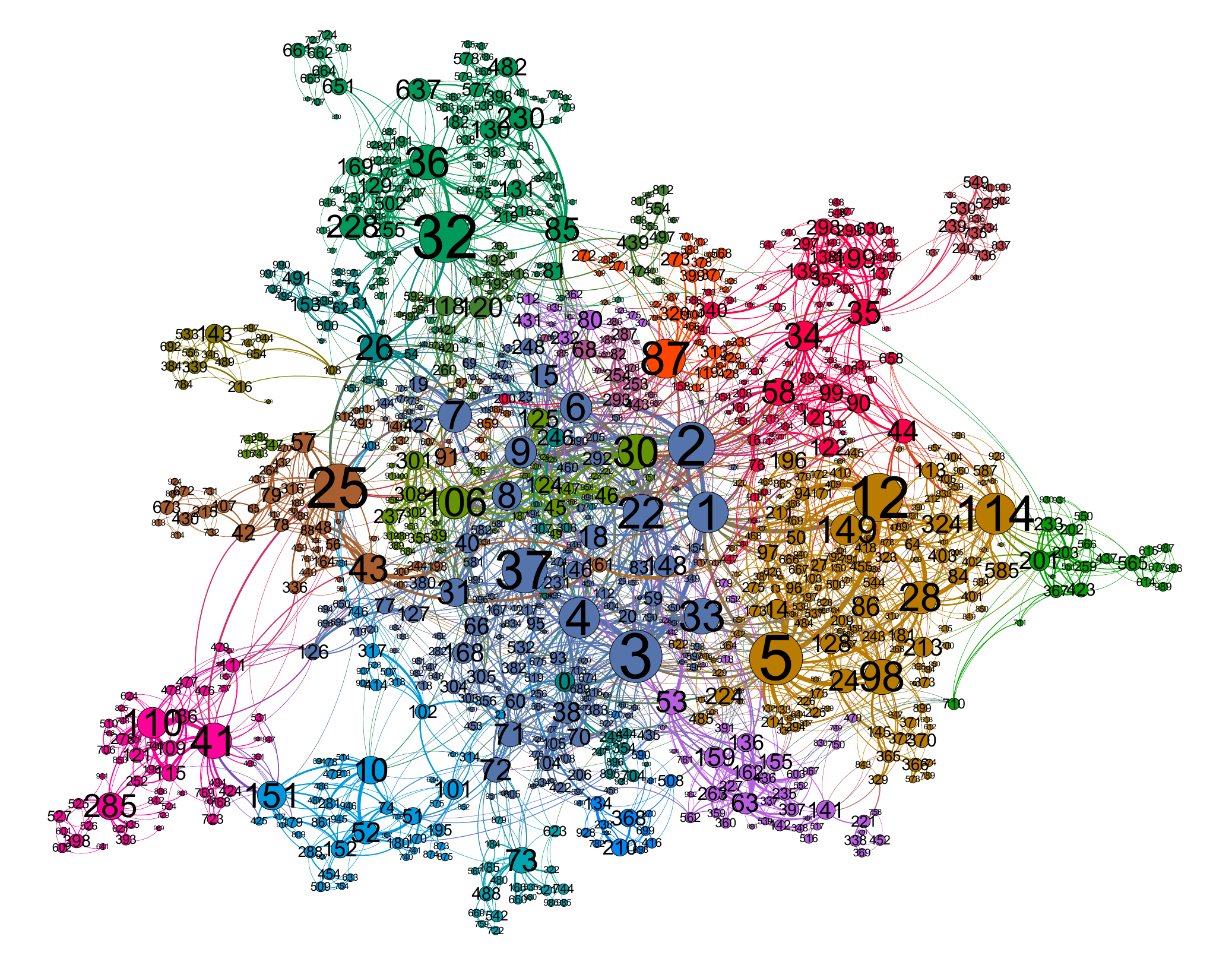}
  \caption{Network: 1000 nodes, $Setting_1$}
	\label{fig:Set1000}
\end{figure}

Figure \ref{fig:Set1000} shows a network with $1000$ nodes and $3599$ edges generated with $Setting_1$. A total of $2830$ interactions took place. The size of nodes and strength of edges correspond to the number of interactions they took part in, the labels indicate the order in which nodes were created. The network has an overlapping community and core-periphery structure. Colored 19 communities were detected by Louvain method, modularity is $0.746$.

\subsection{Evolution of Generated Network}
The subject of the second experiment is one network generated with $Setting_1$. The aim is to show the development of network properties during its growth. The values for each characteristic are measured when the network has $10, 20, 50, 100$, ..., and $10000$ nodes. The results are summarized in Table \ref{tab:Evolut}.

\begin{table*}[ht]
  \centering
  \caption{Evolution of network properties}
	\begin{tabular}{|r|rrrrrrrrrr|}

\hline
  $n$ & $m$ & $<$k$>$ &  $<$l$>$ & $l_{max}$ & $CC$ & $r$ & $com_{IM}$ & $com_L$ & $Q_{IM}$ & $Q_L$ \\
\hline
  10 & 27.00 & 5.4000 & 1.4000 & 20 & 0.8195 & -0.41738 & 1 & 3 & 0.0000 & 0.0590 \\ 
 20 & 53.00 & 5.3000 & 2.0684 & 5 & 0.5804 & -0.09355 & 2 & 4 & 0.0361 & 0.2398 \\ 
  50 & 184.00 & 7.3600 & 2.5004 & 7 & 0.6248 & -0.00886 & 5 & 5 & 0.2957 & 0.3742 \\ 
100 & 388.00 & 7.7600 & 2.8731 & 6 & 0.6793 & 0.09177 & 11 & 9 & 0.4259 & 0.4613 \\ 
 200 & 812.00 & 8.1200 & 3.1859 & 7 & 0.6824 & 0.05107 & 20 & 11 & 0.5002 & 0.5121 \\ 
500 & 2112.00 & 8.4480 & 3.6095 & 8 & 0.6394 & 0.11911 & 44 & 11 & 0.5540 & 0.5849 \\ 
 1000 & 4053.00 & 8.1060 & 3.9402 & 9 & 0.6592 & 0.14528 & 83 & 17 & 0.5812 & 0.6341 \\ 
2000 & 8306.00 & 8.3060 & 4.2054 & 10 & 0.6538 & 0.14016 & 151 & 24 & 0.5972 & 0.6714 \\ 
5000 & 20316.00 & 8.1264 & 4.5344 & 12 & 0.6603 & 0.15823 & 342 & 44 & 0.6203 & 0.6827 \\ 
10000 & 40476.00 & 8.0952 & 4.8015 & 11 & 0.6557 & 0.15351 & 645 & 64 & 0.6307 & 0.7003 \\ 
\hline
  \end{tabular}
	  \label{tab:Evolut}%
\end{table*}%
	
\begin{figure*}[ht]
\centering
\includegraphics[width=46em]{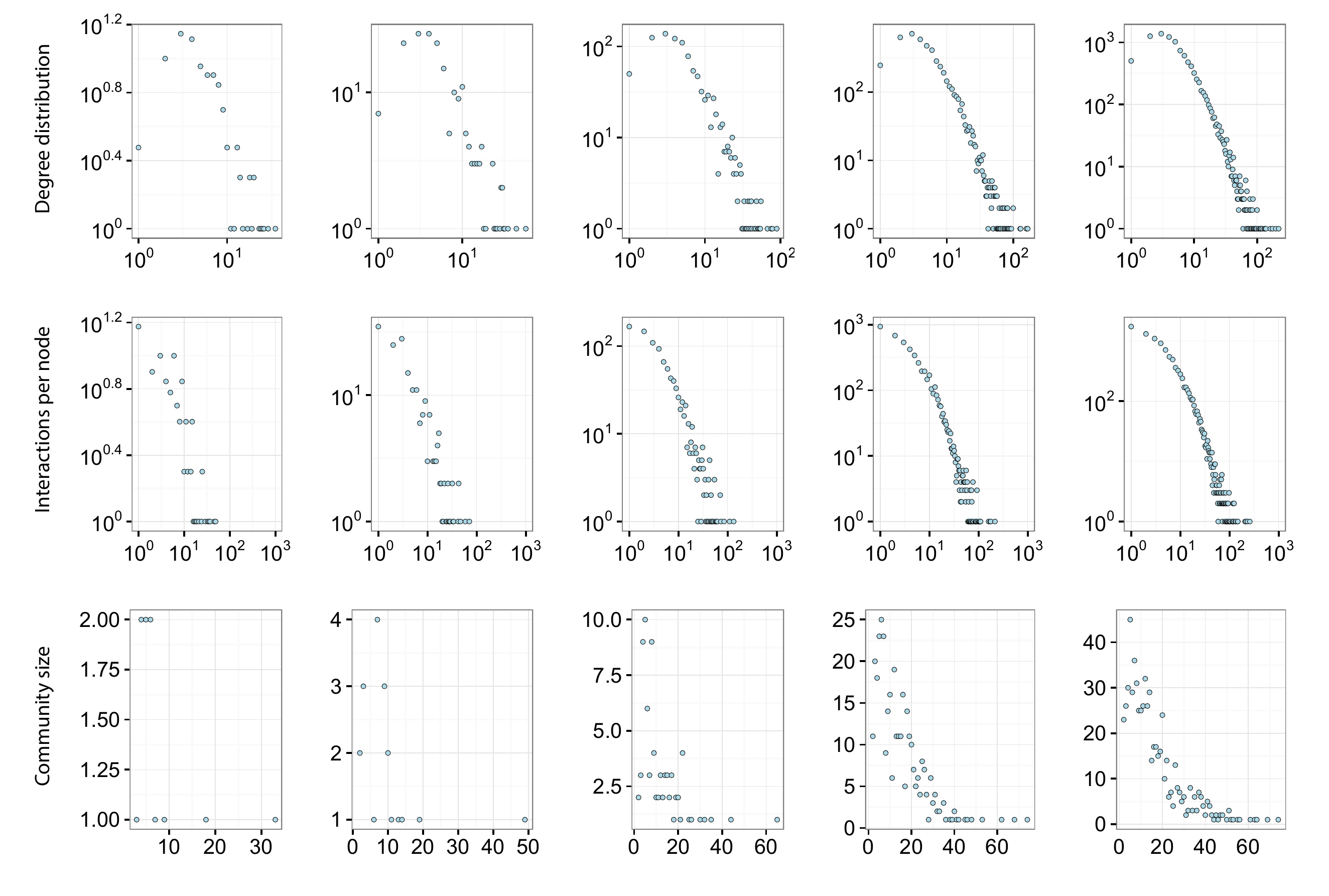}
\caption{Evolution of network (100, 200, 1000, 5000, 10000 nodes)}
\label{fig:EvolNet}
\end{figure*}

Figure \ref{fig:EvolNet}, similarly to the previous experiment, shows the distribution of degree, number of interactions and size of communities. The evolution of these properties is portrayed when network has $100, 200, 1000, 5000, 10000$ nodes.

\begin{figure}[ht]
\centering
\includegraphics[width =\linewidth]{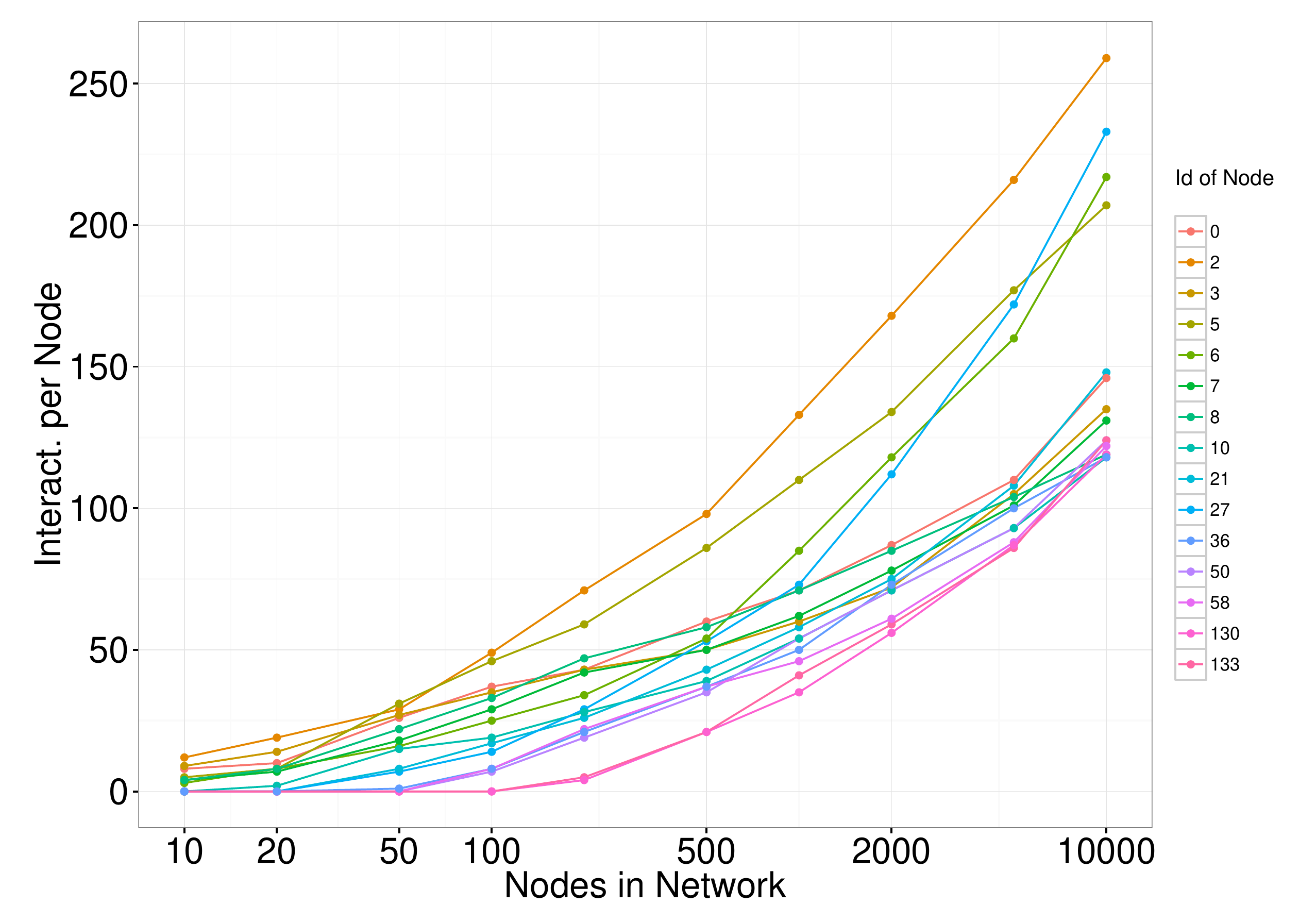}
\caption{Evolution of nodes with the most interactions}
\label{fig:InterEvol}
\end{figure}

Key properties (average degree, shortest path, clustering coefficient, assortativity, modularity) happen to stabilize their values between $1000$ and $10000$ nodes. Our experiments show that  generated networks with other settings also have similar behavior. Figure \ref{fig:InterEvol} shows the evolution of the number of interactions for fifteen nodes with the highest number of interactions at the end of the generation process. The ID of a node represents the moment of its creation. The trend shows how the chances of participating in interactions increase for nodes that already have a high number of interactions.

\subsection{Correlations}
In this experiment, we compare the analyzed real-world network with generated networks. For each setting we generated a network with one million nodes. Then we examined the correlation between the node's creation time and it's degree and number of interactions, respectively. The emergence of the node is represented by its $ID$, which is in the order of its creation. Furthermore, we calculated the correlation between degree and number of node's interactions. We did the same for the analyzed DBLP dataset, where the order of nodes is to be understood as an estimate based on data pre-processing described in Section \ref{sec:dblp}.

\begin{table}[ht]
  \centering
  \caption{Correlations}
\begin{tabular}{|r|rrr|}
	\hline
Setting & $\rho(Id, k)$ & $\rho(Id, i)$ & $\rho(k, i)$ \\ 
\hline
$Setting_1$ & -0.59 & -0.50 & 0.96 \\  
\hline	
$Setting_2$ & -0.47 & -0.47 & 0.95 \\ 
\hline	
$Setting_3$ & -0.70 & -0.58 & 0.87 \\ 
\hline
DBLP & -0.23 & -0.18 & 0.84 \\ 
\hline
	\end{tabular}%
  \label{tab:corr}%
\end{table}%

The results summarized in Table \ref{tab:corr} show that regardless of the setting, the first two correlations are much higher in the 3-lambda model than in the DBLP dataset (Pearson's correlation coefficient was used). Thus, in the presented model, older nodes have a higher (and continuous) chance to participate in interactions than in reality. The cause is probably the aging of nodes in real-world networks where nodes, at different times, no longer participate in interactions. This significantly affects the evolution (growth) and some properties of the network.

Previous experiments demonstrated that despite the absence of aging, generated networks have very good properties. We have not, therefore, for reasons of simplicity, incorporated any of the known models of aging (e.g. inspired by \cite{dorogovtsev2000evolution, xu2010evolutionary}) to the 3-lambda network model.

\section{Conclusion}
\label{sec:conc}
Evolution of real-world networks is influenced by many factors. The purpose of network models is to discover these factors and describe them in a simple way. Our research focused on analyzing behavioral patterns of nodes existing in a co-authorship network while participating in publishing activities. We described four roles of nodes involved in interactions. Based on the analysis of the DBLP dataset, we formulated the hypothesis, which assumes that the numbers of nodes involved in interactions revolve around an average, and they are independent Poisson variables. Based on this hypothesis, we defined the 3-lambda model of collaborative network. The model has three parameters and has no memory. In three experiments, based on three different settings corresponding to dyads, triads and larger groups behavior, we showed that networks generated by the 3-lambda model have the characteristics known from the environment of real-world social networks. Furthermore, we showed that the model can be understood as temporal. In one experiment we presented the development and stabilization of generated network properties in time. For future research there remain some open questions. They bear relation to recognizing other factors influencing the development of the network and to the detailed study of the dependence and predictability of properties of generated networks on the setting of three network parameters (lambdas).

\section{Acknowledgments}
This work was supported by SGS, VSB-Technical University of Ostrava, under the grant ``Parallel Processing of Big Data'', no. SP2016/97.

\bibliographystyle{abbrv}
\bibliography{references}  

\begin{thebibliography}{10}

\bibitem{albert2002statistical}
R.~Albert and A.-L. Barab{\'a}si.
\newblock Statistical mechanics of complex networks.
\newblock {\em Reviews of modern physics}, 74(1):47, 2002.

\bibitem{Arnaboldi2016analysis}
V.~Arnaboldi, R.~I. Dunbar, A.~Passarella, and M.~Conti.
\newblock Analysis of co-authorship ego networks.
\newblock In {\em Advances in Network Science: 12th International Conference
  and School, NetSci-X 2016, Wroclaw, Poland, January 11-13, 2016,
  Proceedings}, volume 9564, page~82. Springer, 2016.

\bibitem{barabasi2002evolution}
A.-L. Barab{\^a}si, H.~Jeong, Z.~N{\'e}da, E.~Ravasz, A.~Schubert, and
  T.~Vicsek.
\newblock Evolution of the social network of scientific collaborations.
\newblock {\em Physica A: Statistical mechanics and its applications},
  311(3):590--614, 2002.

\bibitem{battiston2016emergence}
F.~Battiston, J.~Iacovacci, V.~Nicosia, G.~Bianconi, and V.~Latora.
\newblock Emergence of multiplex communities in collaboration networks.
\newblock {\em PloS one}, 11(1):e0147451, 2016.

\bibitem{bianconi2014triadic}
G.~Bianconi, R.~K. Darst, J.~Iacovacci, and S.~Fortunato.
\newblock Triadic closure as a basic generating mechanism of communities in
  complex networks.
\newblock {\em Physical Review E}, 90(4):042806, 2014.

\bibitem{blondel2008fast}
V.~D. Blondel, J.-L. Guillaume, R.~Lambiotte, and E.~Lefebvre.
\newblock Fast unfolding of communities in large networks.
\newblock {\em Journal of statistical mechanics: theory and experiment},
  2008(10):P10008, 2008.

\bibitem{derenyi2005clique}
I.~Der{\'e}nyi, G.~Palla, and T.~Vicsek.
\newblock Clique percolation in random networks.
\newblock {\em Physical review letters}, 94(16):160202, 2005.

\bibitem{dorogovtsev2000evolution}
S.~N. Dorogovtsev and J.~F.~F. Mendes.
\newblock Evolution of networks with aging of sites.
\newblock {\em Physical Review E}, 62(2):1842, 2000.

\bibitem{evans2010clique}
T.~S. Evans.
\newblock Clique graphs and overlapping communities.
\newblock {\em Journal of Statistical Mechanics: Theory and Experiment},
  2010(12):P12037, 2010.

\bibitem{holme2012temporal}
P.~Holme and J.~Saram{\"a}ki.
\newblock Temporal networks.
\newblock {\em Physics reports}, 519(3):97--125, 2012.

\bibitem{karsai2014time}
M.~Karsai, N.~Perra, and A.~Vespignani.
\newblock Time varying networks and the weakness of strong ties.
\newblock {\em Scientific Reports}, 4:4001, 2014.

\bibitem{leskovec2005graphs}
J.~Leskovec, J.~Kleinberg, and C.~Faloutsos.
\newblock Graphs over time: densification laws, shrinking diameters and
  possible explanations.
\newblock In {\em Proceedings of the eleventh ACM SIGKDD international
  conference on Knowledge discovery in data mining}, pages 177--187. ACM, 2005.

\bibitem{newman2002assortative}
M.~E. Newman.
\newblock Assortative mixing in networks.
\newblock {\em Physical review letters}, 89(20):208701, 2002.

\bibitem{palla2005uncovering}
G.~Palla, I.~Der{\'e}nyi, I.~Farkas, and T.~Vicsek.
\newblock Uncovering the overlapping community structure of complex networks in
  nature and society.
\newblock {\em Nature}, 435(7043):814--818, 2005.

\bibitem{ramasco2004self}
J.~J. Ramasco, S.~N. Dorogovtsev, and R.~Pastor-Satorras.
\newblock Self-organization of collaboration networks.
\newblock {\em Physical review E}, 70(3):036106, 2004.

\bibitem{rombach2014core}
M.~P. Rombach, M.~A. Porter, J.~H. Fowler, and P.~J. Mucha.
\newblock Core-periphery structure in networks.
\newblock {\em SIAM Journal on Applied mathematics}, 74(1):167--190, 2014.

\bibitem{rosvall2009map}
M.~Rosvall, D.~Axelsson, and C.~T. Bergstrom.
\newblock The map equation.
\newblock {\em The European Physical Journal Special Topics}, 178(1):13--23,
  2009.

\bibitem{shekatkar2015complex}
S.~M. Shekatkar and G.~Ambika.
\newblock Complex networks with scale-free nature and hierarchical modularity.
\newblock {\em The European Physical Journal B}, 88(9):1--7, 2015.

\bibitem{song2005self}
C.~Song, S.~Havlin, and H.~A. Makse.
\newblock Self-similarity of complex networks.
\newblock {\em Nature}, 433(7024):392--395, 2005.

\bibitem{toivonen2006model}
R.~Toivonen, J.-P. Onnela, J.~Saram{\"a}ki, J.~Hyv{\"o}nen, and K.~Kaski.
\newblock A model for social networks.
\newblock {\em Physica A: Statistical Mechanics and its Applications},
  371(2):851--860, 2006.

\bibitem{xu2010evolutionary}
K.~S. Xu, M.~Kliger, and A.~O. Hero.
\newblock Evolutionary spectral clustering with adaptive forgetting factor.
\newblock In {\em 2010 IEEE International Conference on Acoustics, Speech and
  Signal Processing}, pages 2174--2177. IEEE, 2010.

\bibitem{yang2012community}
J.~Yang and J.~Leskovec.
\newblock Community-affiliation graph model for overlapping network community
  detection.
\newblock In {\em 2012 IEEE 12th International Conference on Data Mining},
  pages 1170--1175. IEEE, 2012.

\bibitem{zheng2014simple}
B.~Zheng, H.~Wu, L.~Kuang, J.~Qin, W.~Du, J.~Wang, and D.~Li.
\newblock A simple model clarifies the complicated relationships of complex
  networks.
\newblock {\em Scientific reports}, 4, 2014.

\bibitem{zuev2015emergence}
K.~Zuev, M.~Bogu{\~n}{\'a}, G.~Bianconi, and D.~Krioukov.
\newblock Emergence of soft communities from geometric preferential attachment.
\newblock {\em Scientific reports}, 5, 2015.

\end{thebibliography}
%

\balancecolumns 
\end{document}